\documentclass[a4paper]{spie}  %>>> use this instead for A4 paper
\usepackage[]{graphicx}
\usepackage{amssymb,amsmath,psfrag,url}

\newcommand{\Field}[1]{{\boldsymbol{#1}}}
\newcommand{\Kvec}[1]{{\vec{#1}}}

\title{3D Simulations of Electromagnetic Fields in Nanostructures 
using the Time-Harmonic Finite-Element Method} 

\author{
Sven Burger\supit{\,ab}, 
Lin Zschiedrich\supit{\,ab},
Frank Schmidt\supit{\,ab}, 
Roderick K\"ohle\supit{\,c}, \\
Thomas Henkel\supit{\,d}, 
Bernd K\"uchler\supit{\,d}, 
Christoph N\"olscher\supit{\,d}
\skiplinehalf
\supit{a}
Zuse Institute Berlin,
Takustra{\ss}e 7,
D\,--\,14\,195 Berlin,
Germany\\
DFG Forschungszentrum {\sc Matheon},
Stra{\ss}e des 17.\,Juni 136, 
D\,--\,10\,623 Berlin,
Germany
\smallskip\\
\supit{b}
JCMwave GmbH,
Haarer Stra{\ss}e 14a,
D\,--\,85\,640 Putzbrunn, 
Germany
\smallskip\\
\supit{c}
Qimonda AG,
Advanced Technology Software\\
Am Campeon 1-12, D\,--\,85\,579 M\"unchen, Germany
\smallskip\\
\supit{d}
Qimonda Dresden GmbH \& Co.OHG, QD P LM F\\
K\"onigsbr\"ucker Stra{\ss}e 180,
D\,--\,01\,099 Dresden,
Germany
}

\authorinfo{
Corresponding author: S. Burger\\
URL: http://www.zib.de/Numerik/NanoOptics/\\
Email: burger@zib.de
}

\begin{document} 
  \maketitle 
%\today
%%%%%%%%%%%%%%%%%%%%%%%%%%%%%%%%%%%%%%%%%%%%%%%%%%%%%%%%%%%%% 
%%%%%%%%%%%%%%%%%%%%%%%%%%%%%%%%%%%%%%%%%%%%%%%%%%%%%%%%%%%%% 
%% SPIE Copyright form 
\noindent
Copyright 2007  Society of Photo-Optical Instrumentation Engineers.\\
This paper will be published in Proc.~SPIE Vol. {\bf 6617}
(2007),  
({\it Modeling Aspects in Optical Metrology, H.~Bosse, Ed.})
and is made available 
as an electronic preprint with permission of SPIE. 
One print or electronic copy may be made for personal use only. 
Systematic or multiple reproduction, distribution to multiple 
locations via electronic or other means, duplication of any 
material in this paper for a fee or for commercial purposes, 
or modification of the content of the paper are prohibited.
%%%%%%%%%%%%%%%%%%%%%%%%%%%%%%%%%%%%%%%%%%%%%%%%%%%%%%%%%%%%% 
\begin{abstract}

Rigorous computer simulations 
of propagating electromagnetic fields 
have become an important tool for optical metrology 
and optics design of nanostructured components. 
As has been shown in previous benchmarks some of the 
presently used methods suffer from low convergence 
rates and/or low accuracy of the results and exhibit
very long computation times~\cite{Burger2005bacus,Burger2006c}
which makes application to extended 2D layout patterns impractical.
We address 3D simulation tasks by using a finite-element 
solver which has been shown to be superior to competing 
methods by several orders of magnitude in accuracy and 
computational time for typical microlithography 
simulations~\cite{Burger2006c}. 
We report on the current status of 
the solver, incorporating higher order edge elements, 
adaptive refinement methods, and fast solution algorithms.
Further, we investigate the performance of the solver in the 
3D simulation project of light diffraction off an alternating phase-shift 
contact-hole mask.
\end{abstract}

\keywords{3D electromagnetic simulations, microlithography, finite-element methods, FEM}

%%%%%%%%%%%%%%%%%%%%%%%%%%%%%%%%%%%%%%%%%%%%%%%%%%%%%%%%%%%%%

\section{Introduction}

With the advances of micro- and nanotechnology simulation tools 
for rigorous solutions of Maxwell's equations have become an 
important tool in research and development. 
Designing, e.g., a nanooptical component or a metrology tool is 
usually assisted by computer simulations, and simulations are 
used in most scientific works on nanooptical research to 
support theoretical and experimental findings. 
It exists a variety of different methods for solving 
Maxwell's equations, and generally also a variety of 
different numerical implementations of each method. 
Prominent examples of different methods are
the finite-element method (FEM), the finite difference 
time domain method (FDTD), the boundary element method (BEM), 
and rigorously coupled wave analysis (RCWA).

We address 3D simulation tasks occuring in microlithography
by using the frequency-domain finite-element solver {\it JCMsuite}. 
This solver has been successfully applied to a wide 
range of 3D electromagnetic field computations including
microlithography~\cite{Burger2005bacus,Burger2006c,Koehle2007a},
left-handed metamaterials in the optical 
regime~\cite{Linden2006a,Dolling2006a},
photonic crystals~\cite{Burger2006b}, and nearfield-microscopy~\cite{Kalkbrenner2005a}.
The solver has also been used for pattern reconstruction in 
EUV scatterometry~\cite{Pomplun2006bacus,Tezuka2007a}, and it has been benchmarked 
to other methods in typical DUV lithography~\cite{Burger2005bacus,Burger2006c}
and other~\cite{Holzloehner2006a,Burger2006b} projects.

In this paper we report on the current status of the finite-element 
solver {\it JCMsuite}, and we present simulations of light transition through 
periodic arrays of alternating phase-shift contact-holes
which have to be printed on wafers with the same reticle with 
high overlapping process window. 
3D effects cause imbalances of intensities of the $0^\circ$- and 
$180^\circ$-holes, and of different behavior in defocus if the etch depth is not adjusted. 
The evaluation of the impact of these effects has been chosen as a test case for 
{\it JCMsuite}.

\section{Background}

The finite-element package {\it JCMsuite} allows to simulate a variety 
of electromagnetic problems. It incorporates a scattering solver ({\it JCMharmony}), 
a propagating mode solver ({\it JCMmode}) and a resonance solver ({\it JCMresonance}).
The scattering, eigenmode and resonance problems can be formulated 
on 1D, 2D and 3D computational domains. 
Admissible geometries can consist of 
periodic or isolated patterns, or a mixture of both.
Further, solvers for 
problems posed on cylindrically symmetric geometries are implemented. 

\begin{figure}[htb]
\centering
\psfrag{0d}{\sffamily $0^\circ$}
\psfrag{1d}{\sffamily $180^\circ$}
\psfrag{px}{\sffamily $\mbox{p}_{\mbox{x}}$}
\psfrag{py}{\sffamily $\mbox{p}_{\mbox{y}}$}
\includegraphics[width=0.5\textwidth]{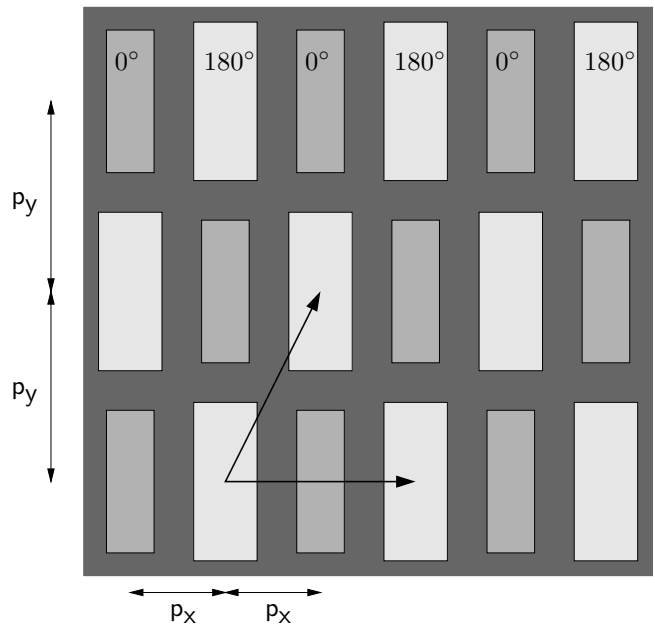}
\caption{
Schematics of the 3D test structure: cross-section in a $x$-$y$-plane. 
0-degree (180-degree) rectangular phase-shift holes are depicted
in grey (light grey). The elementary vectors, $\vec{a}_1, \vec{a}_2$, of the periodic pattern are 
indicated by arrows. 
}
\label{schema_geo_1}
\end{figure}

\begin{figure}[htb]
\centering
\psfrag{p_x}{\sffamily $\mbox{p}_{\mbox{x}}$}
\psfrag{0d}{\sffamily $0^\circ$}
\psfrag{1d}{\sffamily $180^\circ$}
\psfrag{d3}{\sffamily $\mbox{d}_{\mbox{Cr,top}}$}
\psfrag{d2}{\sffamily $\mbox{d}_{\mbox{Cr,bottom}}$}
\psfrag{d1}{\sffamily $\mbox{d}_{\mbox{MoSi}}$}
\psfrag{detch}{\sffamily $\mbox{d}_{\mbox{etch}}$}
\psfrag{substrate}{\sffamily substrate}
\psfrag{superspace}{\sffamily superspace (air)}
\includegraphics[width=0.5\textwidth]{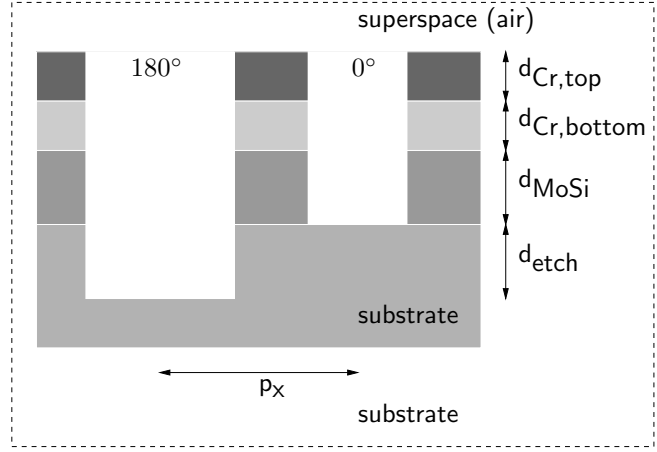}
\caption{
Schematics of the 3D test structure: cross-section in a $x$-$z$-plane. 
The material stack on a Quartz (SiO$_2$) substrate consists of a molydenum silicide
and two layers containing chromium. 
0-degree (180-degree) rectangular phase-shift air-filled holes are indicated.
}
\label{schema_geo_2}
\end{figure}

\begin{figure}[htb]
\centering
\psfrag{cdx0}{\sffamily $\mbox{CD}_{\mbox{x,0}}$}
\psfrag{cdx1}{\sffamily $\mbox{CD}_{\mbox{x,180}}$}
\psfrag{cdy0}{\sffamily $\mbox{CD}_{\mbox{y,0}}$}
\psfrag{cdy1}{\sffamily $\mbox{CD}_{\mbox{y,180}}$}
\psfrag{comp_domain}{\sffamily computational domain}
\includegraphics[width=0.5\textwidth]{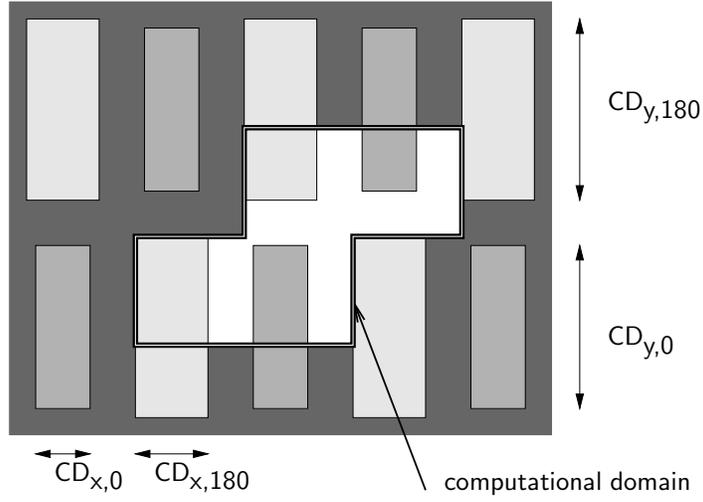}
\caption{
Schematics of the 3D test structure: cross-section in a $x$-$z$-plane. 
The computational domain is indicated by a polygon.
}
\label{schema_geo_3}
\end{figure}

In this paper we concentrate on 
light-scattering  off a 3D pattern (photomask) which is 
periodic in the $x-$ and $y-$directions and is enclosed by  homogeneous 
substrate (at $z_{sub}$) and superstrate (at $z_{sup}$) 
which are infinite in the $-$, resp.~$+z-$direction (see Figures~\ref{schema_geo_1}
and~\ref{schema_geo_2}). 
Light propagation in the investigated system is governed by Maxwell's equations
where  vanishing densities of free charges and currents are assumed~\cite{Wong2005a}. 
The dielectric coefficient $\varepsilon(\vec{x})$ and the permeability 
$\mu(\vec{x})$ of the considered photomasks are periodic and complex, 
$\varepsilon \left(\vec{x}\right)  =  \varepsilon \left(\vec{x}+\vec{a} \right)$, 
$\mu \left(\vec{x} \right)  =  \mu \left(\vec{x}+\vec{a} \right)$.
Here $\vec{a}$ is any elementary vector of the periodic lattice.  
For given primitive lattice vectors 
$\vec{a}_{1}$ and $\vec{a}_{2}$ an elementary cell 
$\Omega\subset\mathbb R^{3}$ can be defined as
$\Omega = \left\{\vec{x} \in \mathbb R^{2}\,|\,
x=\alpha_{1}\vec{a}_1+\alpha_{2}\vec{a}_2;
0\leq\alpha_{1},\alpha_{2}<1
\right\}
\times [z_{sub},z_{sup}]$.
As can be seen from the computational domain  in 
Figure~\ref{schema_geo_3} elementary cells  $\Omega$ of the same volume 
can be defined also 
by more complicated polygons which are adapted to the actual geometry.
\footnote{In this case a computational domain constructed corresponding to 
$\Omega = \left\{\vec{x} \in \mathbb R^{2}\,|\,
x=\alpha_{1}\vec{a}_1+\alpha_{2}\vec{a}_2;
0\leq\alpha_{1},\alpha_{2}<1
\right\}
\times [z_{sub},z_{sup}]$
would in general lead to intersections of the domain boundary and geometrical features 
at sharp angles. This could result in low quality triangulations.
}

A time-harmonic ansatz with frequency $\omega$ and magnetic field 
$\Field{H}(\vec{x},t)=e^{-i\omega t}\Field{H}(\vec{x})$ leads to
the following equations for $\Field{H}(\vec{x})$:
\begin{itemize}
\item
The wave equation and the divergence condition for the magnetic field:
\begin{eqnarray}
\label{waveequationH}
\nabla\times\frac{1}{\varepsilon(\vec{x})}\,\nabla\times\Field{H}(\vec{x})
- \omega^2 \mu(\vec{x})\Field{H}(\vec{x}) &=& 0,
\qquad\vec{x}\in\Omega,\\
\label{divconditionH}
\nabla\cdot\mu(\vec{x})\Field{H}(\vec{x}) &=& 0,
\qquad\vec{x}\in\Omega .
\end{eqnarray}
\item
Transparent boundary conditions at the boundaries to the 
substrate (at $z_{sub}$) and superstrate (at $z_{sup}$), $\partial\Omega$,
where $\Field{H}^{in}$ is the incident magnetic field (plane wave 
in this case), and $\vec{n}$ is the normal vector on $\partial\Omega$:
\begin{equation}
\label{tbcH}
	\left(
        \frac{1}{\varepsilon(\vec{x})}\nabla \times (\Field{H} - 
        \Field{H}^{in})
	\right)
	\times \vec{n} = DtN(\Field{H} - 
        \Field{H}^{in}), \qquad \vec{x}\in \partial\Omega.
\end{equation}
The $DtN$ operator (Dirichlet-to-Neumann) is  realized with 
an adaptive PML method~\cite{Zschiedrich03,Zschiedrich2006a}. 
This is a generalized formulation of Sommerfeld's radiation condition.
%; it
%can be realized alternatively by the Pole condition method~\cite{Hohage03a}.
\item
Periodic boundary conditions for the transverse boundaries, $\partial\Omega$,
governed by Bloch's theorem~\cite{Sakoda2001a}:
\begin{equation}
\label{bloch}
\Field{H}(\vec{x}) = e^{i \Kvec{k}\cdot\vec{x}} \Field{u}(\vec{x}), \qquad
\Field{u}(\vec{x})=\Field{u}(\vec{x}+\vec{a}),
\end{equation}
where the Bloch wavevector $\Kvec{k}\in\mathbb{R}^3$ is defined by the
incoming plane wave $\Field{H}^{in}$.

\end{itemize}

Similar equations are found for the electric field 
$\Field{E}(\vec{x},t)=e^{-i\omega t}\Field{E}(\vec{x})$;
these are treated accordingly.
The finite-element method solves Eqs.~(\ref{waveequationH}) -- (\ref{bloch})
in their weak form, i.e., in an integral representation. 

The finite-element methods consists of the following steps:
\begin{itemize}
\item
The computational domain is discretized with simple geometrical patches,
{\it JCMsuite} uses linear (1D), triangular (2D) and tetrahedral or prismatoidal 
(3D) patches. 
The use of prismatoidal patches is well suited for layered geometries, as in 
photomask simulations. 
%Sidewall angles different from 90\,deg are not regarded throughout this paper;
%however, they can easily be implemented with reasonable restrictions. 
\item
The function spaces in the integral representation of Maxwell's equations 
are discretized using Nedelec's edge elements, 
which are vectorial functions of polynomial order 
defined on the simple geometrical patches~\cite{Monk2003a}. 
In the current implementation, {\it JCMsuite} uses polynomials of 
first ($1^{st}$) to ninth ($9^{th}$) order. 
In a nutshell, FEM can be explained as expanding the field 
corresponding to the exact solution of Equation~(\ref{waveequationH}) in the 
basis given by these elements.
\item
This expansion leads to  a large sparse matrix equation (algebraic problem).
To solve the algebraic problem on a standard workstation 
linear algebra decomposition techniques (LU-factorization, e.g.,
package PARDISO~\cite{PARDISO}, which was used in the simulations of 
Chapter~\ref{3dchapter})
%or iterative and domain decomposition 
%methods~\cite{Zschiedrich2005b} are used, 
%depending on problem size.
are used. 
In cases with either large computational domains or high accuracy 
demands, also domain decomposition methods~\cite{Zschiedrich2005b} 
are used and allow to handle problems with very large numbers of unknowns.
\end{itemize}

For details on the weak formulation, 
the choice of Bloch-periodic functional spaces,
the FEM discretization, and the implementation of the adaptive PML
method in {\it JCMsuite}
we refer to previous works~\cite{Zschiedrich03,Zschiedrich2006a}.
In future implementations performance will further be increased 
by using curvilinear elements, general domain-decomposition techniques  and 
$hp$-adaptive methods.

\section{3D simulations of a contact-hole photomask}
\label{3dchapter}
\subsection{Parameter settings and simulation flow}
We apply the 3D light-scattering solver of the programme package
{\it JCMsuite}  to the 
test case of a 2D periodic pattern of an alternating phase-shift
contact-hole photomask. 
Figures~\ref{schema_geo_1},~\ref{schema_geo_2}  and~\ref{schema_geo_3} 
show cross-sections 
through the geometrical layout of the mask and define the geometrical parameters. 
The geometrical parameters investigated in this study are given 
in Table~\ref{table_apsm}.
The computational domain is a unit-cell of the periodic pattern; it is indicated 
in Figure~\ref{schema_geo_3}.
Its total volume is $V=2 p_x \times p_y \times h_z$, where 
$h_z = d_{\mbox{Cr,top}}+{d}_{\mbox{Cr,bottom}}+{d}_{\mbox{MoSi}}+{d}_{\mbox{etch}}$.
Please note that a simple rectangular box-like computational domain would have 
twice the volume, $\tilde{V}=2 p_x \times 2 p_y \times h_z$.
Within the numerical error the results are independent of the choice of the unit-cell 
chosen as computational 
domain.

\begin{figure}[htb]
\centering
\psfrag{Layout}{\sffamily Geometry layout}
\psfrag{Triangulation}{\sffamily Mesh parameters}
\psfrag{Materials}{\sffamily Material parameters}
\psfrag{Mesh}{\sffamily Mesh}
\psfrag{Sources}{\sffamily Sources parameters}
\psfrag{Project}{\sffamily Project parameters}
\psfrag{Results}{\sffamily Results}
\psfrag{Projection}{\sffamily Projection}
\psfrag{Interface}{\sffamily MATLAB Interface: Automatic input generation,
parameter scans, data analysis}
\includegraphics[width=0.85\textwidth]{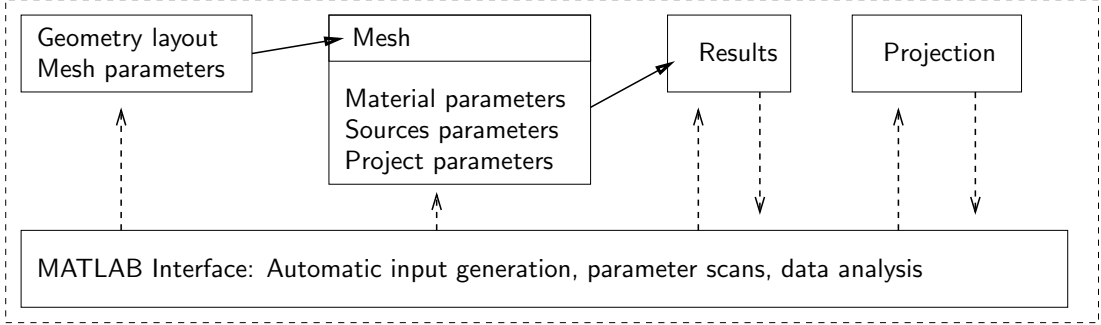}
\caption{
Schematics of the simulation flow of the FEM solver 
{\it JCMsuite}.
}
\label{schema_simulation_flow}
\end{figure}

\begin{table}[h]
\begin{center}
\begin{tabular}{|l|l|l|l|l|l|l|}
\hline
parameter & set 1&  set 2&  set 3&  set 4&  set 5&  set 6\\ 
\hline 
\hline 
$p_x$ [nm] &  576& 640& 720& 800& 1040& 1200 \\ 
$p_y$ [nm] & 1120& 1120& 1120& 1120& 1120& 1120  \\ 
$\mbox{CD}_{\mbox{x,0}}$ [nm]  & 416& 452& 496& 552& 496& 520\\
$\mbox{CD}_{\mbox{y,0}}$ [nm] & 876&   800&   752&   696&   976&   980\\
\hline 
$\mbox{CD}_{\mbox{x,180}}$ [nm] & \multicolumn{6}{l|} {$\mbox{CD}_{\mbox{x,0}} + \mbox{3D bias}$}\\
$\mbox{CD}_{\mbox{y,180}}$ [nm] & \multicolumn{6}{l|} {$\mbox{CD}_{\mbox{y,0}} + \mbox{3D bias}$}\\
3D bias [nm] & \multicolumn{6}{l|}{(32, 48, 64, 80)}\\
$\mbox{d}_{\mbox{Cr,top}}$ [nm] & \multicolumn{6}{l|} {20}\\
$\mbox{d}_{\mbox{Cr,bottom}}$ [nm] & \multicolumn{6}{l|}{40}\\
$\mbox{d}_{\mbox{MoSi}}$ [nm] & \multicolumn{6}{l|}{68}\\
$\mbox{d}_{\mbox{etch}}$ [nm] & \multicolumn{6}{l|}{(156, 160, 164, 168, 172, 176, 180)}\\
\hline 
\hline 
$\lambda_0$ & \multicolumn{6}{l|} {193.0\,nm } \\ 
$\varepsilon_{r \mbox{air}}$ & \multicolumn{6}{l|} {1.0}\\
$\varepsilon_{r \mbox{Cr,top}}$ & \multicolumn{6}{l|} {$(1.871 + 1.13i)^2$}\\
$\varepsilon_{r \mbox{Cr,bottom}}$ & \multicolumn{6}{l|}{$(1.477 + 1.762i)^2$}\\
$\varepsilon_{r \mbox{MoSi}}$ & \multicolumn{6}{l|}{$(2.343 + 0.586i)^2$}\\
$\varepsilon_{r \mbox{SiO}_2}$ & \multicolumn{6}{l|} {$(1.564)^2$}\\
\hline 
\end{tabular} 
\caption{Parameter settings for the 3D simulations.
The relative permittivity, $\varepsilon_r$, is the square of the complex 
index of refraction, $\varepsilon_r=n^2=(n_r+i k)^2$.
The simulation wavelength was $\lambda=193.0\,$nm while the wavelength 
in the real system is $\lambda=(193.36 \pm 0.2)\,$nm. 
The dispersion of $\varepsilon$ is taken into account at $\lambda=193.36\,$nm. 
All combinations of 3D bias and etch depth, $\mbox{d}_{\mbox{etch}}$, 
have been investigated for the six
parameter sets in order to find for each pattern the best value and the 
best overall 3D bias and etch depth.
}
\label{table_apsm}
\end{center}
\end{table}

\begin{figure}[h!]
\centering
\psfrag{(a)}{\sffamily (a)}
\psfrag{(b)}{\sffamily (b)}
\psfrag{(c)}{\sffamily (c)}
\includegraphics[width=0.8\textwidth]{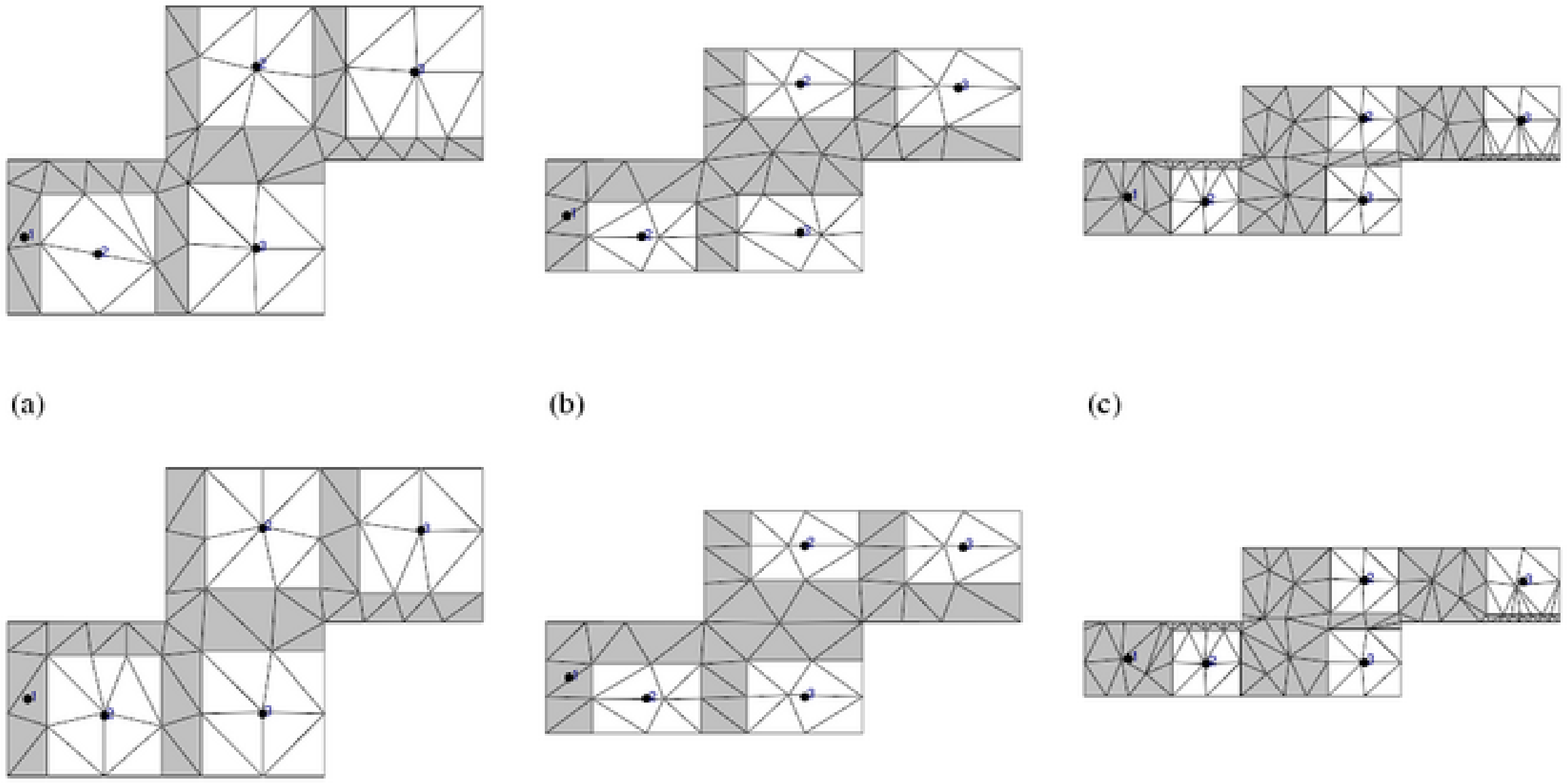}
\caption{
Cross-sections through prismatoidal discretizations of several 
computational domains with different geometry 
parameters, corresponding to parameter sets 1 (a), 4 (b), 6 (c)(see Table~\ref{table_apsm}).
Please note that (a), (b), and (c) have different length scales.
}
\label{triangulations}
\end{figure}

Figure~\ref{schema_simulation_flow} shows the simulation flow of the 
finite-element software:
\begin{itemize}
\item
The geometry of the computational domain is described in a polygonal 
format of the cross-section, corresponding to Fig.~\ref{schema_geo_1}, and 
includes a height profile (see Fig.~\ref{schema_geo_2}) with  material
attributions and statements about the computational domain boundaries 
(periodic and transparent boundaries in this case).
The translational vectors of the periodic pattern ($\vec{a}_1, \vec{a}_2$)
are identified automatically from the layout, optimized settings 
of the perfectly matched layers (PML) are found automatically with an 
adaptive method~\cite{} (adaptive PML, aPML).
Further, parameters specifying a maximum patch size of the 
finite-element mesh and further meshing properties can be set.
From these geometry parameters and mesh parameters the prismatoidal mesh is 
generated automatically. Figure~\ref{triangulations} shows $(x-y)$-cross-sections
through the prismatoidal meshes for different geometrical parameter sets. 
\item
Material parameters (complex permittivity and permeability tensors) 
can be specified as piecewise 
constant functions and/or (using dynamically loaded libraries)  
as functions of arbitrary spatial dependence. 
In the presented example, the piecewise constant, isotropic settings given 
in Table~\ref{table_apsm} are used. The relative permeability is $\mu_r=1$
for all used materials.
\item
Light sources can be defined as predefined functions (plane waves, Gaussian 
beams, point sources) or as arbitrary functions using dynamically loaded libraries. 
{\it JCMsuite} allows to generate solutions to several independent source terms 
in parallel, efficiently re-using the inverted system matrix. 
In the presented example, two plane waves with normal incidence (wavevector 
$\vec{k}=(0,0,k_z)$), and with orthogonal polarizations are used. 
\item
The main project definitions in this case are the accuracy settings (mesh 
refinement, PML refinement, finite-element degree) and the definitions of 
postprocesses.
Upon execution the {\it JCMsolve} computes the full field distribution over the entire 
computational domain. 
Through internal or external post-processes, the quantities of interest can be derived from the 
field. E.g., the complex amplitudes of propagating modes are attained 
by Fourier transformation, an aerial image is calculated~\cite{Koehle2007a}, 
or the field distribution is exported to graphics format for 
visualisation and analysis.
\item
Interfaces to scripting languages like Python or MATLAB can be used for 
performing automatic parameter scans and data analysis.
In the presented example, the MATLAB interface for the generation of 
the input files and for the execution of a loop over the 
geometrical parameters given in Table~\ref{table_apsm} consists of 
well below 100 lines of MATLAB code.
\end{itemize}
 
\subsection{Error analysis}

\begin{figure}[htb]
\centering
\psfrag{(a)}{\sffamily (a)}
\psfrag{(b)}{\sffamily (b)}
%\psfrag{N}{\sffamily N [\,$10^5$\,]}
%\psfrag{approx. total time [min]}{\sffamily total time [\,min\,]}
\includegraphics[width=0.95\textwidth]{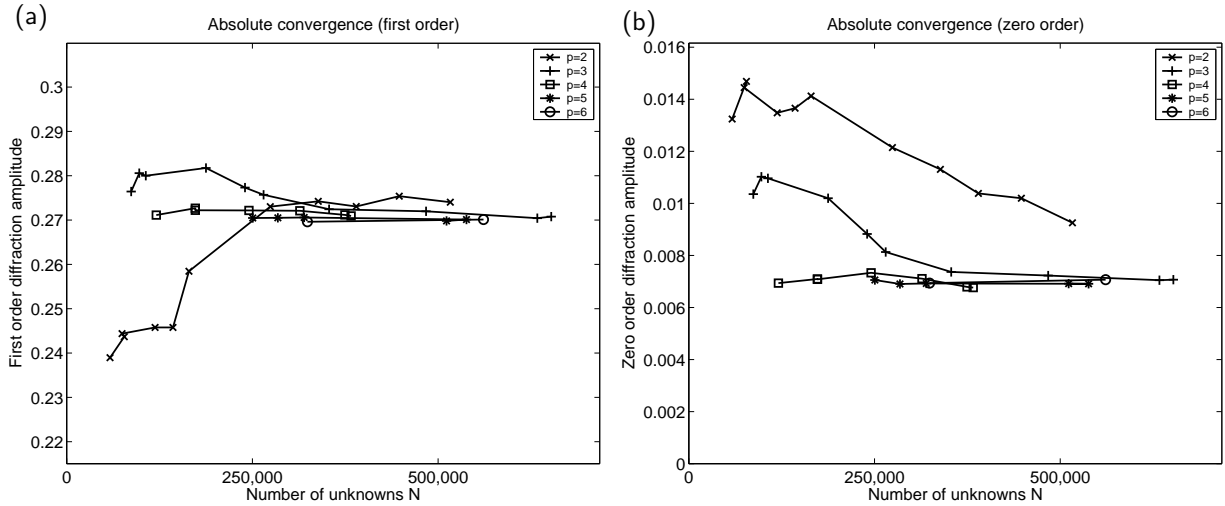}
\caption{
Convergence of the intensities of the first (a) and zero (b) diffraction orders
with increasing number of unknowns of the FEM solution. 
Results from computations with finite element degree $p=2\dots 6$ are 
displayed.
}
\label{convergence_1a}
\end{figure}

\begin{figure}[htb]
\centering
\psfrag{(a)}{\sffamily (a)}
\psfrag{(b)}{\sffamily (b)}
%\psfrag{N}{\sffamily N [\,$10^5$\,]}
%\psfrag{approx. total time [min]}{\sffamily total time [\,min\,]}
\includegraphics[width=0.95\textwidth]{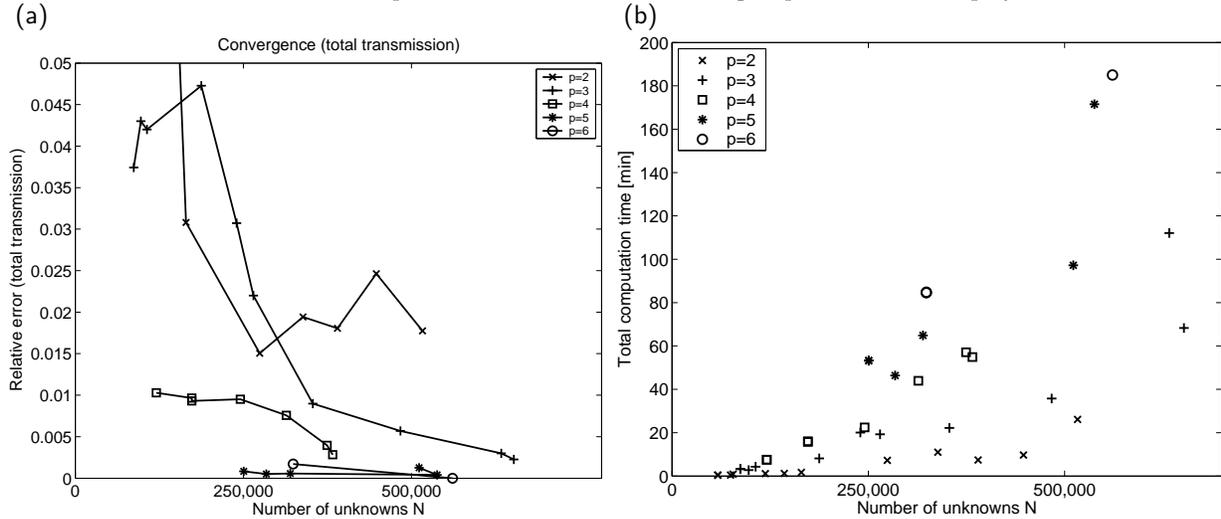}
\caption{
(a) Convergence of the total transmission amplitude to the 5 x 5 lowest 
diffraction orders.
(b) Computation times in dependence of the number of unknowns of the corresponding 
matrix equation.
}
\label{convergence_1b}
\end{figure}

We have investigated the accuracy of the near field results by 
simulating near fields using finite-elements of different polynomial degrees 
($p$-refinement)
and using meshes with different numbers of mesh elements ($h$-refinement)
for one fixed setting of geometrical and other physical parameters.
From each near fields we compute the diffraction pattern using a post-process. 
Figure~\ref{convergence_1a} shows the convergence of the absolute values of the 
intensities in the zero and first diffraction orders.
Figure~\ref{convergence_1b}(a) shows the convergence of the relative error of the intensity 
$I$ in the 5 x 5 lowest diffraction orders.
The relative error of the intensity, $\Delta I_r$, is defined as 
$(I-I_{qe})/I_{qe}$, where the quasi-exact intensity, $I_{qe}$,
is deduced from a solution with highest finite-element degree and finest 
mesh. 
As can be seen from Figure~\ref{convergence_1b}, relative errors of below 
one percent for the prominent diffraction orders can be reached using 
a high polynomial degree of the finite-elements ($p\ge 4$) and relatively 
moderate numbers of unknowns, corresponding to coarse spatial discretizations. 

\subsection{Computational costs}

Figure~\ref{convergence_1b}(b) shows the computation time in minutes in dependence 
on the number of degrees of freedom in the simulation. Different symbols in the 
plot correspond to different finite-element polynomial degrees. 
The computations have been performed on a standard workstation with 16
(2 x 8)  processors and extended RAM (AMD Opteron, 64GB RAM). 
Please note that for the results in Figure~\ref{convergence_1b}(b) we 
have used only four of the 16 processors, while for the scans over geometrical 
parameters we have 
used twelve out of the 16 processors, giving a speed-up factor of approximately 
three in the computation times.
As can be seen in Figure~\ref{convergence_1a} and
Figure~\ref{convergence_1b}, results corresponding to an accuracy of the 
prominent diffraction orders of better 1\% 
can be attained in computation times of few minutes when using finite 
elements of order $p=4$ and above. 
Memory requirements were roughly 5\dots 15GB RAM for $N\sim 2.5\dots 5\cdot 10^5$
and finite-element degree $p=4$ (the setting used for the scan over the geometrical 
parameters).
Please note that in general it does depend on the physical problem 
wheather a refinement of the finite element degree $p$, $p\rightarrow p+1$,
or a refinement of the mesh width $h$, $h\rightarrow h/2$, leads to a 
better convergence of the results.
For the given rather large (on the wavelength-scale) structures on the 
photomask, rather high degrees $p$ seem to lead to the best results, while 
for problems with smaller features finer meshes are favourable. 
In a future implementation  of the programme package 
this choice will be automatically and locally detected by $hp$-adaptive methods.

\begin{figure}[htb]
\centering
\psfrag{(a)}{\sffamily (a)}
\psfrag{(b)}{\sffamily (b)}
%\psfrag{N}{\sffamily N [\,$10^5$\,]}
%\psfrag{approx. total time [min]}{\sffamily total time [\,min\,]}
\includegraphics[width=0.5\textwidth]{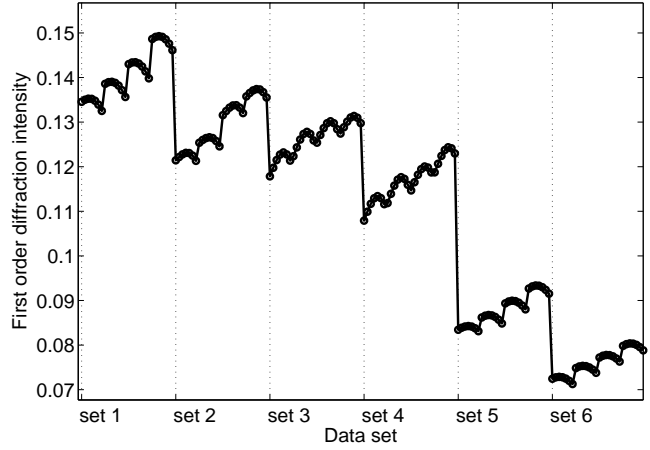}
\caption{
Intensity of the first (+1,0) diffraction order  for the different investigated parameter sets. 
Each parameter set consists of four groups (3D bias=8,12,16,20\,nm)  of seven 
values of etch depths ($\mbox{d}_{\mbox{etch}} =156, 160, 164, 168, 172, 176, 180$\,nm).
}
\label{first_order_diffraction}
\end{figure}

\begin{figure}[htb]
\centering
\psfrag{N}{\sffamily N [\,$10^5$\,]}
\psfrag{approx. total time [min]}{\sffamily total time [\,min\,]}
\includegraphics[width=0.5\textwidth]{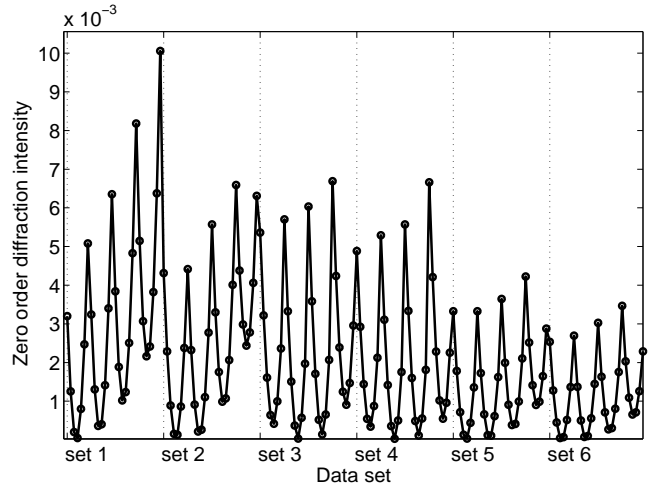}
\caption{
Intensity of the zero (0,0) diffraction order  for the different investigated parameter sets. 
Each parameter set consists of four groups (3D bias=8,12,16,20\,nm)  of seven 
values of etch depths ($\mbox{d}_{\mbox{etch}} =156, 160, 164, 168, 172, 176, 180$\,nm).
}
\label{zero_order_diffraction}
\end{figure}

\subsection{Geometry dependence of the diffraction spectrum}

We have performed a scan over the geometrical parameters defined in 
Table~\ref{table_apsm}. The six parameter sets consist each of 28 subsets of 
the various combinations of 3D bias and etch depth. 
Figure~\ref{first_order_diffraction} and~\ref{zero_order_diffraction} 
show how the first  (+1,0) order 
and zero order diffraction intensities vary over the scan of geometrical 
parameters. 
As expected for phase-shift masks, the intensity in the first diffraction orders
(prominent order) is about 
two orders of magnitude higher than the intensity in the zero diffraction order. 
For otherwise fixed geometrical parameters, the variation of the prominent diffraction 
order with the etch depth in the investigated regime is about 2-4\%. 
The variation of the prominent diffraction order with the 3D bias is about 10-15\%. 
The variation with a different set of CD and pitch parameters is much larger. 

This shows that for mask design using rigorous simulations, especially for 
the design of parameters such as the etch depth,  it is necessary to use 
simulation tools which produce results of low numerical error.
% (lower than 2\% in this case).

\begin{figure}[htb]
\centering
\psfrag{(a)}{\sffamily (a)}
\psfrag{(b)}{\sffamily (b)}
\includegraphics[width=0.95\textwidth]{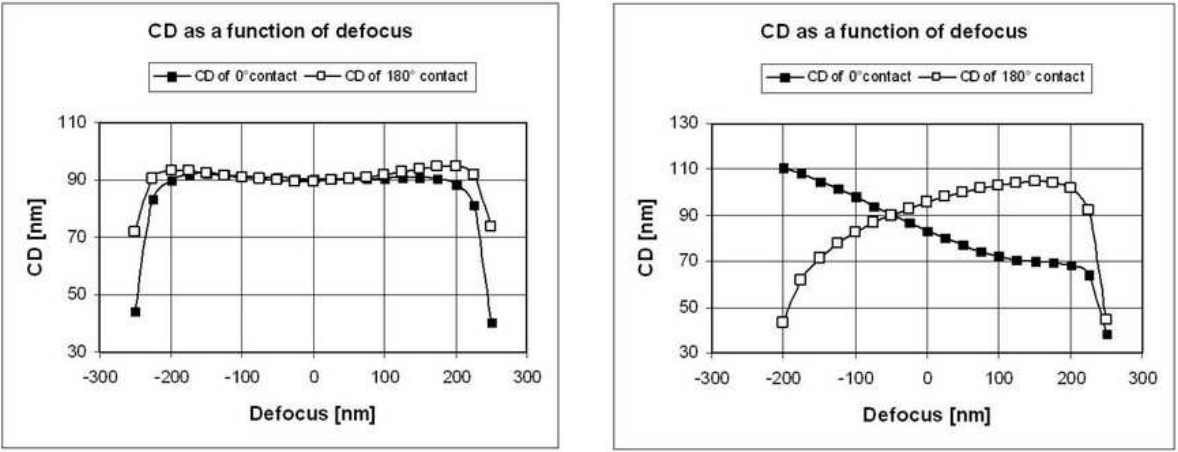}
\caption{
(a) CD as a function of defocus for an almost balanced case.
(b) CD as a function of defocus for an imbalanced case.
}
\label{qifig1}
\end{figure}
\subsection{Imaging application}

For the conditions of Table~\ref{table_apsm} the 
farfield coefficients were determined and used as input for an 
Qimonda inhouse imaging simulator. 
The mask bias for each pitch was pre-optimized by 
Kirchhoff (2D) mask simulations in resist.
For evaluation of the arial image output for 
each pitch and 3D bias the intensity threshold for target CD was determined 
at best focus as average of 0$^\circ$- and 180$^\circ$-holes.
Figure~\ref{qifig1}(b) shows the typical phase balancing problem that 
occurs in case of non-optimum etch depth in the mask: 
the CD of 0$^\circ$- and 180$^\circ$-holes behaves different at defocus, 
whereas for the optimum depth the behavior is symmetric (Figure~\ref{qifig1}(a)).

\begin{figure}[htb]
\centering
\psfrag{(a)}{\sffamily (a)}
\psfrag{(b)}{\sffamily (b)}
\includegraphics[width=0.95\textwidth]{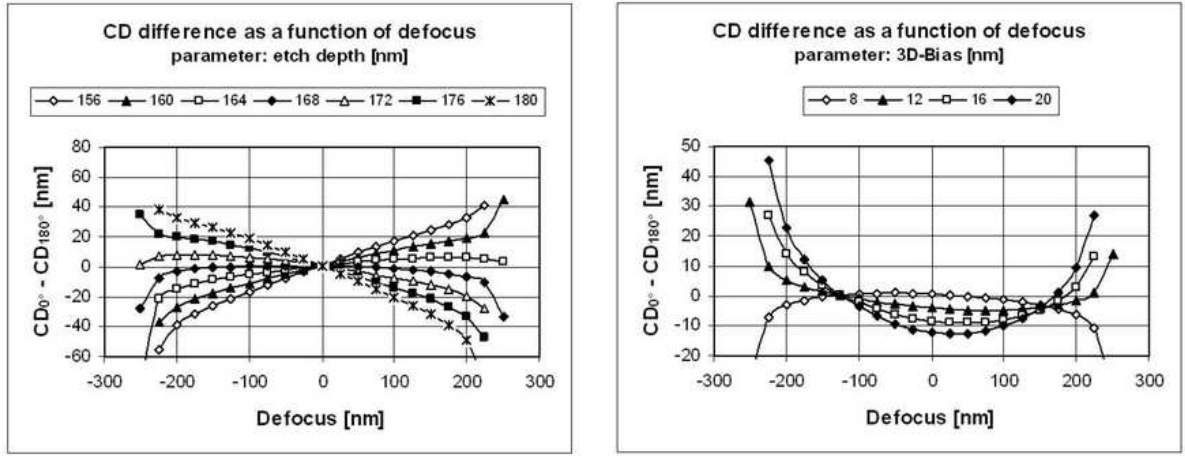}
\caption{
(a) CD difference ($\mbox{CD}_{\mbox{0}}-\mbox{CD}_{\mbox{180}}$) 
as a function of defocus and etch depth for one fixed 3D bias.
(b) CD difference ($\mbox{CD}_{\mbox{0}}-\mbox{CD}_{\mbox{180}}$) 
as a function of defocus and 3D bias at optimized etch depth.
}
\label{qifig2}
\end{figure}

\begin{figure}[htb]
\centering
\psfrag{(a)}{\sffamily (a)}
\psfrag{(b)}{\sffamily (b)}
\includegraphics[width=0.95\textwidth]{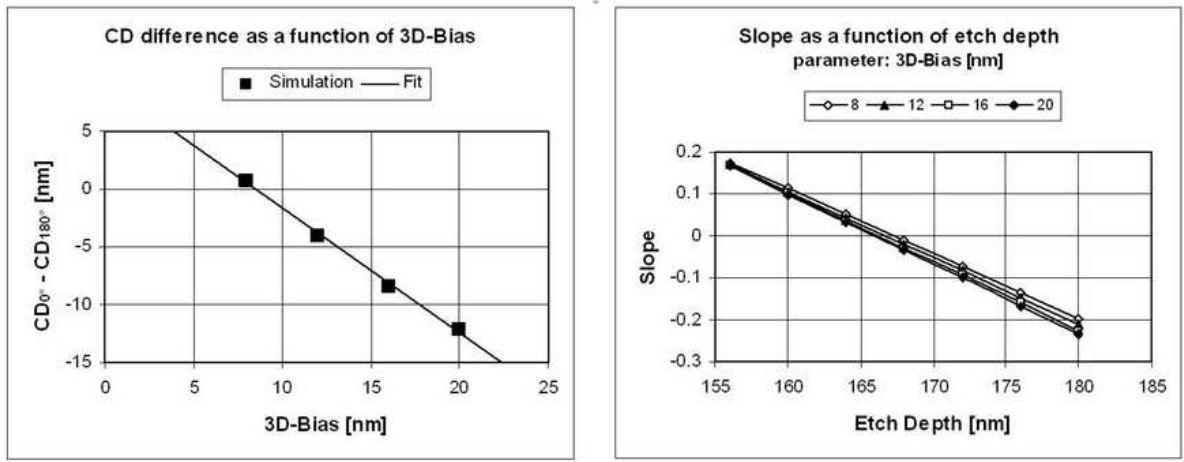}
\caption{
(a) CD difference ($\mbox{CD}_{\mbox{0}}-\mbox{CD}_{\mbox{180}}$) 
as a function of 3D bias at optimized etch depth (168\,nm).
(b) Slope of  CD difference,  
$\partial(\mbox{CD}_{\mbox{0}}-\mbox{CD}_{\mbox{180}})/\partial (\mbox{defocus})$, 
as a function of etch depth and 3D bias.
}
\label{qifig3}
\end{figure}

The CD difference of 0$^\circ$- and 180$^\circ$-holes 
over defocus shows a slope that is characteristic for the phase error, i.e.,
the deviation from optimum etch depth where the defocus behavior is flat 
(see Figure~\ref{qifig2}(a)). 
Furtheron, the 180$^\circ$-hole prints smaller than a 0$^\circ$-hole of same 
size due to 3D mask effects. 
This is compensated by a 3D mask bias, see Figures~\ref{qifig2}(b) and~\ref{qifig3}(a),  
which is in the range of 8\,--\,20\,nm, depending on pattern and mask conditions 
(side wall angles, material). 
The slope is rather independent of the 3D bias, as shown in Figure~\ref{qifig3}(b).
The optimum 3D bias, where the CD difference is zero, is 8.5\,nm at an etch 
depth of 168\,nm, as obtained by a linear fit of the CD difference vs.~3D bias. 
As can be also seen in Figure~\ref{qifig3}(b), the slope is almost zero at 
this etch depth. 

A determination of the optimum etch depths and optimum 3D mask 
biases through pitch reveals significant differences. 
While the 3D mask bias can be corrected individually for such special patterns by
software algorithms it is challenging to correct the 3D mask 
effect in real layouts containing random patterns.
An even more critical issue is to find an optimum etch depth 
through pitch. Here an analysis of overlapping
process windows including etch depth deviations is the next 
step to find an optimum etch depth through pitch.
Then conclusions can be drawn with respect to lithographic 
applications for nodes below 40\,nm half pitch.

\section{Conclusions}
\label{conclusions}
We have performed rigorous 3D FEM simulations of light transition through 
alternating phase-shift contact-hole photomasks using the finite-element 
programme package {\it JCMsuite}. 
We have investigated the convergence behavior of the solutions and we have 
shown that we achieve results at high numerical accuracy. 
Our results show that rigorous 3D mask simulations can well 
be handled at high accuracy and relatively low computational cost.

The investigation of the diffraction spectrum of an alternating phase-shift 
contact-hole mask has shown that for mask design and optimization projects 
it is necessary to use rigorous tools with low discretization errors,
typically below 1\% for the prominent diffraction orders,
because the variance of the diffraction orders can be of the order of few percent
for typical design parameter regimes. 
We have used the simulations to determine optimum etch depths and 3D mask 
biases for the various investigated parameter sets.

\acknowledgements
This paper was supported by the EFRE fund of the European Community 
and by funding of the State Saxony of the Federal Republic of Germany (project number
10834). 
The authors are responsible for the content of the paper.

%\bibliography{./phcbibli,./group05,./lithography}   
\bibliography{/home/numerik/bzfburge/texte/biblios/phcbibli,/home/numerik/bzfburge/texte/biblios/group,/home/numerik/bzfburge/texte/biblios/lithography}   
%\bibliography{/home/numerik/bzfburge/texte/biblios/phcbibli,/home/numerik/bzfburge/texte/biblios/NanoOptics,/home/numerik/bzfburge/texte/biblios/lithography}   
\bibliographystyle{spiebib}   

\end{document}